\documentclass[twocolumn,preprintnumbers,amsmath,amssymb,prb,superscriptaddress]{revtex4}
\usepackage{color,epsfig,rotating,graphicx,subfigure}

\usepackage[latin1]{inputenc}
\usepackage[english]{babel}
\usepackage{dsfont}
\usepackage{textcomp}

\renewcommand{\Im}{{\rm Im}}
\newcommand{\ri}{{\rm i}}
\newcommand{\re}{{\rm e}}

\newcommand{\Tr}{{\rm Tr}}

\graphicspath{{./Figures/}}

\begin{document}

\title{Limitations of the kinetic theory to describe the\\near-field heat exchanges in many-body systems}

\date{\today}

\author{Christoph Kathmann}
\affiliation{Institut f{\"u}r Physik, Carl von Ossietzky Universit{\"a}t, D-26111 Oldenburg, Germany}
\author{Riccardo Messina}
\email{riccardo.messina@institutoptique.fr}
\affiliation{Laboratoire Charles Fabry, UMR 8501, Institut d'Optique, CNRS, Universit\'{e} Paris-Saclay, 2 Avenue Augustin Fresnel, 91127 Palaiseau Cedex, France}
\author{Philippe Ben-Abdallah}
\email{pba@institutoptique.fr}
\affiliation{Laboratoire Charles Fabry, UMR 8501, Institut d'Optique, CNRS, Universit\'{e} Paris-Saclay, 2 Avenue Augustin Fresnel, 91127 Palaiseau Cedex, France}
\author{Svend-Age Biehs}
\email{s.age.biehs@uni-oldenburg.de}
\affiliation{Institut f{\"u}r Physik, Carl von Ossietzky Universit{\"a}t, D-26111 Oldenburg, Germany}

\begin{abstract}
We investigate the radiative heat transfer along a chain of nanoparticles using both a purely kinetic approach based on the solution of a Boltzmann transport equation and an exact method (Landauer's approach) based on fluctuational electrodynamics. We show that the kinetic theory generally fails to predict properly the heat flux transported along the chain both at close (near-field regime) and large separation (far-field regime) distances. We report a deviation of a factor two between the heat fluxes predicted by the two approaches in the diffusive regime of heat transport and we show that this difference becomes even greater than two orders of magnitude in the ballistic regime.
\end{abstract}

\maketitle

\section{Introduction}

The study of radiative heat transfers between two bodies separated by a large gap is an old problem in physics. It goes back to the beginning of the development of the blackbody theory~\cite{Kirchoff,Planck} and it is at the origin of the development of quantum physics. During the 70's these pioneering works have been extended by Polder and van Hove~\cite{Polder} to describe also the radiative heat exchange between two bodies at separation distances smaller than the thermal wavelength, where evanescent waves participate to the heat flux, by photon tunneling and surface mode coupling. 

In the 2000's a first attempt of generalization of these works to a more general scenario involving $N$ bodies in mutual interaction has been performed~\cite{1,2} in the case of simple nanoparticle networks at zero temperature through which heat is carried by a classical multiscattering process. But contrary to Polder and van Hove's theoretical framework which is based on the fluctuational-electrodynamics theory~\cite{Rytov} here a kinetic approach has been adopted to deal with the heat transfer. This theory is based on the solution of a Botzmann transport equation for the distribution function of thermal photons. Hence the heat flux inside these systems results from the calculation of first order moments associated to the distribution function. Moreover, it is intrinsically related to the presence of resonant modes supported by the structure. Hence, only the eigenstates of system are assumed to play a role in the heat transport process. Recently various complex systems have been investigated using such an approach and unexpected thermal behaviors~\cite{3,5,6,4} have been predicted.

Beside this approximate theory a rigorous strategy has been followed in 2011 to describe the near-field heat transfers in $N$-body systems consisting of small size body~\cite{9,10,8,Edalatpour2014,11,12,EkerothEtAl2017} and later to describe more general situations with systems of arbitrary size and of various geometries~\cite{MA14,14} by generalizing Rytov's theory to many-body systems allowing also to take into account the spatial distribution of temperature profiles~\cite{16}. Following these theoretical developments numerous many-body effects have been highlighted in these systems. For example, anomalous heat transports regimes have been demonstrated~\cite{13,17}, $N$-body amplification mechanisms~\cite{MessinaPRL13} or magneto-optical effects~\cite{PBA16}, paving the way to new functionalities for thermal management at nanoscale~\cite{Transistor,Memory,Boolean}.

\begin{figure}[h!]
 \includegraphics[width=0.45\textwidth]{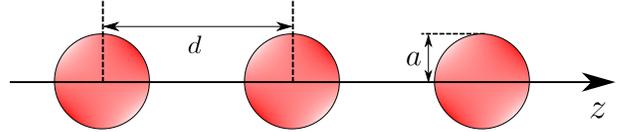}
\caption{Sketch of the considered configuration: a set of $N$ spherical nanoparticles aligned along the $z$ axis. The particle have all radius $a$, while $d$ is the center-to-center distance between neighboring particles.}
\label{Fig:SketchGeometry}\end{figure}

In this work we question the limits of the kinetic approach to describe the radiative heat exchange in $N$-body systems. To address this fundamental question of the validity of the Boltmann approach we consider here a simple chain of spherical nanoparticles and compare the predictions of the Boltzmann approach with those of the exact Landauer approach. The paper is organized as follows. In Sec.~\ref{SecTheory} we remind the general expression of heat flux and thermal conductance using both theoretical frameworks, by distinguishing in the Boltzmann formalism the ballistic and the diffusive regimes of heat transport. Then in Sec.~\ref{SecModes} we study the resonant modes (eigenstates) supported by the system. We derive their dispersion relations and their propagation length inside the structure as well. In Sec.~\ref{SecComparison} we effectively compare the heat flux and heat flux spectra predicted by the Landauer and Boltzmann approach, while in Sec.~\ref{SecDiscussion} we discuss a variety of configurations in which the disagreement highlighted here can be even more striking and we give some conclusive remarks.

\section{Radiative heat transfer in a chain of nanoparticles}\label{SecTheory}

The system we consider in the following is a set of $N$ identical spherical nanoparticles with radius $a$ and center-to-center distance $d$ aligned along the $z$ axis as sketched in Fig. \ref{Fig:SketchGeometry}. The calculations we are going to present are based on the dipolar approximation, according to which each particle is described as a dipolar pointlike emitter. This approximation is valid as long as the radius $a$ of the nanoparticles is small compared to all the other relevant lengthscales in the system. We choose here $a=25\,$nm, much smaller than the wavelengths relevant to the radiative heat transfer (in the micron range), and we limit our calculations to distances satisfying $d>3a$. While the dipolar approximation is safely valid for $d>4a$, we also present results in the range $3a<d<4a$, in which the dipolar term is still expected to be a relevant contribution to the energy transfer. In the dipolar approximation the optical response of each particle is described in terms of the electrical polarizability, which in the quasi-static approximation has a Clausius-Mossotti-like form an reads
\begin{equation}
 \alpha(\omega)= 4 \pi a^3\frac{\epsilon(\omega)-1}{\epsilon(\omega)+2}.
\end{equation}
It depends on the permittivity $\epsilon(\omega)$ of the material. In the following, we will consider particles made of hexagonal boron nitride (hBN), for which we will employ the Drude-Model $\epsilon(\omega)=\varepsilon_\infty(\omega^2-\omega_L^2+i\Gamma\omega)/(\omega^2-\omega_R^2+i\Gamma\omega)$ with parameters $\varepsilon_\infty=4.88$, $\omega_L=3.032\times10^{14}\,\text{rad s}^{-1}$, $\omega_R=2.575\times10^{14}\,\text{rad s}^{-1}$ and $\Gamma=1.001\times10^{12}\,\text{rad s}^{-1}$\,[\onlinecite{NarayanaswamyApplPhysLett03,spitzer}]. This model predicts the existence of a surface phonon-polariton resonance at the material-vacuum spherical interface [corresponding to a zero of $\varepsilon(\omega)+2$] at frequency $\omega_0\simeq2.906\times10^{14}\,\text{rad s}^{-1}$.

We want to calculate the radiative heat transfer within the chain of nanoparticles when a given temperature difference is imposed, and compare the results obtained with the Landauer and Boltzmann approaches. In the following we will consider a chain made of $N=100$ nanoparticles, and assume that the temperatures of the particles at the boundaries are fixed at the values $T_1=350\,$K and $T_N=300\,$K by contact with two external reservoirs. Before presenting the numerical results, we will briefly present in the following Sections the main assumptions and equations associated with the two approaches.

\subsection{Landauer approach}

The Landauer approach is an exact theoretical framework based on fluctuational electrodynamics. The main tool of this theory is the fluctuation-dissipation theorem, which connects the statistical properties of the field generated by each body in the system to its optical properties, i.e. to its dielectric permittivity (in the case of non-magnetic particles considered in this work). In our specific scenario, simplified by the dipolar approximation, each particle is described in terms of a fluctuating electric dipole $\mathbf{p}_j$ ($j=,1,\dots,N$), with a zero average value and correlations given by
\begin{equation}\begin{split}
 \langle p_{j,\beta}^{\rm (fl)}(\omega)p_{j',\beta'}^{{\rm (fl)}\dagger}(\omega')\rangle & = \hbar \epsilon_0 \delta_{jj'} \delta_{\beta\beta'} \chi_j(\omega) 2\pi \delta(\omega-\omega')\\
 &\,\times[1+2n(\omega,T_j)].
 \label{fldis}
\end{split}\end{equation}
In this expression the Latin indices refer to the different particles, while the Greek ones refer to the Cartesian coordinates. Moreoever, the susceptibility $\chi_j$ of each particle is defined as~\cite{abajo}
\begin{equation}
 \chi_j=\Im(\alpha_j)-\frac{\omega^3}{6\pi c^3} |\alpha_j|^2,
\end{equation}
and 
\begin{equation}
 n(\omega,T) = \biggl[\exp\biggl(\frac{\hbar\omega}{k_B T}\biggr) - 1\biggr]^{-1}.
\end{equation}
is the Bose--Einstein distribution at temperature $T$. The information concerning the statistical properties of the field emitted by the individual dipoles is completed, in this approach, by the rigorous solution of the scattering problem in the $N$-body system which can be conveniently expressed, for instance, by means of the Green function of the system. The detailed derivation in a system of $N$ dipoles is presented in Ref.~\onlinecite{10}. The final result for the power absorbed by particle $i$ reads
\begin{equation}P_i=\int_0^{+\infty}\frac{d\omega}{2\pi}\hbar\omega\sum_{j\neq i}\frac{4\chi_i\chi_j}{|\alpha_i|^2}n_{ji}(\omega)\Tr\Bigl(\mathds{T}^{-1}_{ij}\mathds{T}^{-1\dag}_{ji}\Bigr),\end{equation}
where we have introduced the differences
\begin{equation}n_{ij}(\omega)=n(\omega,T_i)-n(\omega,T_j),\end{equation}
and $\mathbb{T}$ is a $3N\times3N$ block matrix defined in terms of the $(i,j)$ $N\times N$ sub-matrices ($i,j=1,\dots,N$)
\begin{equation}
 \mathds{T}_{ij} = \delta_{ij}\mathds{1}-(1-\delta_{ij})\frac{\omega^2}{c^2}\alpha_i\mathds{G}_{ij},
\end{equation}
$\mathds{G}_{ij}$ being the Green function in vacuum evaluated at the coordinates of dipoles $i$ and $j$.

\subsection{Boltzmann approach}

As anticipated, the exact results obtained with the Landauer approach will be compared to the ones derived using the kinetic Boltzmann approach. This framework is based on the Boltzmann transport equation, and we will present it in two opposite regimes, namely the diffusive and the ballistic regimes. The definition of these two regimes is based on the comparison between the total length of the chain and the characteristic propagation length of the resonant phonon-polariton modes existing within the chain, to which the transport of heat is attributed.

\subsubsection{Diffusive regime}

In the diffusive regime the chain length is assumed to be much larger than the propagation length of the localized phonon-polaritons along the chain. As a result, phonon-polaritons existing within the chain undergo a large number of scattering events. This results in a temperature profile along the chain. The difference in the local Bose-Einstein distributions finally leads to a heat flux along the chain, which can for a chain of length $L=dN$ and diameter $S=\pi a^2$ be described by the Boltzmann transport equation as~\cite{2}
\begin{equation}
\begin{split}
 j&= \frac{1}{LS} \sum_{k=-\infty}^\infty \hbar \omega_k f_k v_{g,k}\\
 &= \frac{1}{LS} \sum_{k=-\infty}^\infty \hbar \omega_k [f_k-n(\omega_k,T)] v_{g,k},
\end{split}
\label{jdicht}
\end{equation}
where $v_{g,k}$ is the group velocity of the mode $k$, and $f_k$ the distribution function. The latter is given by the solution of the Boltzmann equation~\cite{haken}
\begin{equation}
 \frac{\partial f}{\partial t}+v_{g,k}\frac{\partial f}{\partial z}=\biggl[\frac{\partial f}{\partial t}\biggr]_{\rm coll}.
 \label{boltz}
\end{equation}
The first term vanishes in the stationary regime. The second term can be approximated by the simplified expression 
\begin{equation}
 v_{g,k}\frac{\partial f}{\partial z}=v_{g,k} \frac{\partial n(\omega,T(z))}{\partial T}\frac{dT}{dz}
\end{equation}
assuming the system close to the thermal equilibrium along the chain.
Moreover, in the relaxation time approximation The righthand side of Eq.~\eqref{boltz} can be simplified into
\begin{equation}
 \biggl[\frac{\partial f}{\partial t}\biggr]_{\rm coll}=-\frac{f-n(\omega,T)}{\tau_{k}},
\end{equation}
where $\tau_k$ denotes the relaxation time toward the equilibrium. 
With these simplifications the deviation of the distribution from equilibrium can be expressed as
\begin{equation}
 f-n(\omega,T)=-\tau_{k}v_{g,k}\frac{\partial n(\omega,T(z))}{\partial T}\frac{dT}{dz}.
\end{equation}
Inserting this in Eq.~\eqref{jdicht} and using $\Lambda_k=|\tau_k v_{g,k}|$ results in
\begin{equation}
 j= \frac{1}{LS} \sum_{k} \hbar \omega_k \biggl[-\Lambda_k \frac{\partial n(\omega,T(x))}{\partial T}\frac{dT}{dz}\biggr] v_{g,k}.
\end{equation}
The summation over $k$ can be replaced by an integration over $\omega$ using the density of states and the definition of the group velocity $v_g=\frac{d\omega}{dk}$. When perfmorming this replacement, one has to put as boundaries of the frequency integration the minimum and maximum values of the dispersion relation $\omega_\text{min}$ and $\omega_\text{max}$, respectively, which depend on the polarization (see e.g. Fig.~\ref{fig:disps}), the radius $a$ of the particles and the distance $d$ between them. The new expression reads
\begin{equation}
\begin{split}
 j&=-\frac{1}{\pi S} \int_{\omega_\text{min}}^{\omega_\text{max}} \!\!{d\omega}\, \hbar \omega \Lambda_\omega \frac{\partial n(\omega,T(x))}{\partial T}\frac{dT}{dz}.\\
\end{split}
\end{equation}
The heat flux expression has the form of Fourier's law of heat conduction
\begin{equation}
 j=-\kappa \frac{dT}{dz},
\end{equation}
where
\begin{equation}
 \kappa=\frac{1}{\pi S} \int_{\omega_\text{min}}^{\omega_\text{max}} {d\omega} \hbar \omega \Lambda_\omega \frac{\partial n(\omega,T(x))}{\partial T}
\end{equation}
is the thermal conductivity. From this expression the heat flux $P_d$ along a chain of length $L$ with its ends being coupled to heat baths at different temperatures can be calculated. Assuming that the temperature difference between the ends is much smaller than their average, one obtains
\begin{equation}
\begin{split}
 P_d &=\frac{S}{L}\kappa \Delta T\\
 &=\frac{1}{\pi}\int_{\omega_\text{min}}^{\omega_\text{max}} \!\!{d\omega} \, \hbar \omega \frac{\Lambda_\omega}{L} \frac{\partial n(\omega,T(x))}{\partial T} \Delta T\\
 &=\int_{\omega_\text{min}}^{\omega_\text{max}}\frac{d\omega}{2\pi} \, \hbar \omega n_{1N} \frac{2\Lambda_\omega}{L},
\end{split}
\end{equation}
This is exactly the same expression as derived in Ref.~\onlinecite{3}. To obtain the total heat flux, this expression needs to be evaluated for both the two transversal and the single longitudinal coupled modes and then all contributions have to be added. Therefore, we obtain the final expression
\begin{equation}\label{DiffFin}
 P_d=\int_{\omega_\text{min}}^{\omega_\text{max}}\frac{d\omega}{2\pi} \, \hbar \omega n_{1N} \frac{2}{L} (2\Lambda_\omega^{\perp}+\Lambda_\omega^{\parallel}).
\end{equation}

\subsubsection{Ballistic regime}

In the ballistic regime, the chain length is assumed to be much smaller than the propagation length of the localized phonon-polaritons along the chain. Therefore, phonon-polaritons can propagate along the chain without being scattered. As a result, every mode in the dispersion relation fully contributes to the heat transport. The net heat flux is the difference of the fluxes emitted by the heat baths at the ends of the chain
\begin{equation}
\begin{split}
 j&=j_{l\rightarrow r}+j_{r\rightarrow l}\\
 &= \frac{1}{LS} \sum_{k>0} \hbar \omega_k [n(\omega,T_1) -n(\omega,T_N)] v_{g,k}.
\end{split}
\end{equation}
As before the sum over $k$ can be replaced by an integration over $\omega$ using the density of states to obtain the heat flux per cross sectional area
\begin{equation}
 j= \frac{1}{S}\int_{\omega_\text{min}}^{\omega_\text{max}} \frac{d\omega}{2\pi} \hbar \omega [n(\omega,T_1) -n(\omega,T_N)].
\end{equation}
Multiplication with the cross sectional area of the chain and summation over all polarizations results in the total heat flux
\begin{equation}
 P_b=3\int_{\omega_\text{min}}^{\omega_\text{max}} \frac{d\omega}{2\pi} \hbar \omega\, n_{1N}.
\label{Eq:ballisticflux}
\end{equation}
This expression has also been evaluated in Ref.~\onlinecite{3}, for instance. It gives the largest possible heat flux along the chain, because each mode will fully contribute to the energy transfer. Note, that both transversal and the longitudinal coupled modes contribute in the same way which explains the prefactor $3$. 

\section{Resonant modes}\label{SecModes}

In the previous Section, we have seen that the group velocity $v_{g,k}$ and the relaxation time $\tau_k$ play a crucial role in the determination of the propagation length $\Lambda_\omega$, based on which the heat flux in the diffusive regime can be easily calculated using Eq.~\eqref{DiffFin}. These two quantities can be deduced from the dispersion relation of the modes propagating along the chain of particles. To this aim we start from the expression of the electric field emitted by a single dipole $\mathbf{p}$ located at $\mathbf{r}'$, which reads
\begin{equation}
 \textbf{E}(\textbf{r})=\omega^2 \mu_0 \mathbb{G}(\textbf{r},\textbf{r}')\cdot \textbf{p},
\end{equation}
As there is no external field in our system, the dipole moment of the $n$-th particle induced by all other particles is given by
\begin{equation}
 \textbf{p}_n = \frac{\omega^2}{c^2} \alpha(\omega) \sum_{m \neq n} \mathbb{G}_{nm} \cdot \textbf{p}_m,
\end{equation}
where $\mathbb{G}_{nm}=\mathbb{G}(\textbf{r}_n,\textbf{r}_m)$ is Green function at position $\textbf{r}_n$ of the $n$-th dipole due to the $m$-th dipole $\textbf{p}_m$ at position $\textbf{r}_m$. We now assume that we have a strictly periodic chain (thus, with an infinite number of particles), and consider a travelling wave along the chain. This allows us to replace the $n$-th dipole by $\textbf{p}_n= \textbf{p}_0 \re^{\ri nkd}$, resulting in a dispersion relation for modes propagating along the chain for both polarizations. For the transverse polarization we obtain~\cite{ford}
\begin{equation}
\begin{split}
 0 &=1+2\frac{\alpha(\omega)}{4 \pi d^3}\sum_{j=1}^\infty \biggl[ \biggl( 1- \ri\frac{\omega d}{c}j \biggr) \frac{1}{j^3} \\ 
 &\qquad\qquad\qquad\qquad-\frac{\omega^2 d^2}{c^2}\frac{1}{j} \biggr] \cos(jkd) \re^{\ri \frac{\omega d}{c} j},
\end{split}\label{dispt}\end{equation} 
and for the longitudinal polarization we find
\begin{equation}
0 =1-4\frac{\alpha(\omega)}{4 \pi d^3}\sum_{j=1}^\infty \biggl[ \biggl( 1-\ri\frac{\omega d}{c}j \biggl) \frac{1}{j^3} \biggr] \cos(jkd) \re^{\ri \frac{\omega d}{c} j}.
\label{displ}\end{equation}
These equations need to be solved in the complex plane. In the following we will work assuming a real wavevector $k$ as an independent variable, and findind a complex $\omega=\omega'+i\omega''$ from Eqs.~\eqref{dispt} and \eqref{displ}. We remark that the factor $\re^{\ri \frac{\omega d}{c} j}$ physically imposes $\Im(\omega) \geq 0$ in order to have a mode whose amplitude vanishes for $t\to\infty$. A possible simplified approach to obtain the dispersion relation is to work in the quasistatic limit, assuming $c \rightarrow \infty$. In this case the Eqs.~\eqref{dispt} and \eqref{displ} simplify considerably and the solution can be found for real $k$ and $\omega$. Once the dispersion relation has been obtained, the group velocity is simply given by $d\omega'/dk$, whereas the relaxation time is $1/2\omega''$. We conclude that the propagation length appearing in Eq.~\eqref{DiffFin} reads
\begin{equation}\label{Lambda}
 \Lambda_\omega = \frac{d\omega'}{dk}\frac{1}{2\omega''}.
\end{equation}

\begin{figure}[t]
 \includegraphics[width = 0.48\textwidth]{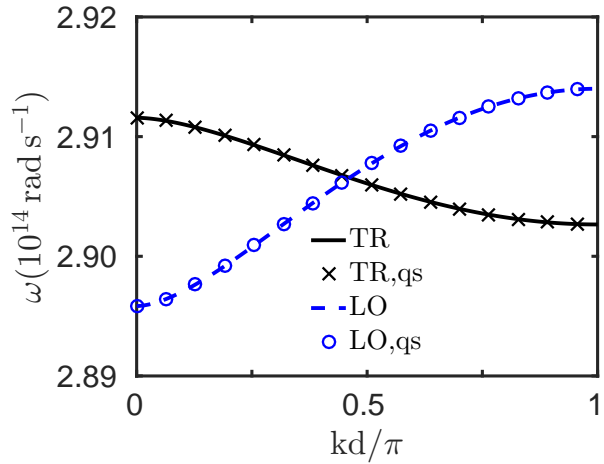}
\caption{Dispersion relation for a chain of hBN nanoparticles with radius $a=25\,$nm and distance $d=4a=100\,$nm for transverse (black curves) and longitudinal (blue curves) polarizations. The lines correpond to the search of complex frequency $\omega$ (the plot shows the real part of $\omega$) using real $k$, while the symbols correspond to the quasistatic limit.}
\label{fig:disps}
\end{figure}

\subsection{Dispersion relations}

We now present the numerical solutions of Eqs.~\eqref{dispt} and \eqref{displ} in the case of an infinite chain of hBN nanoparticles of radius $a=25\,$nm and distance $d=4a=100\,$nm. We show in Fig.~\ref{fig:disps} the real part $\omega'$ of the frequency as a function of the wavevector $k$ for both polarizations. These results are compared to the quasistatic approach. As a result of the periodicity of the system, the problem can be solved in the First Brillouin Zone (FBZ), i.e. for $0<k<\pi/d$.

We clearly see that the modes within the chain exist in a specific range of frequencies, which depends both on the considered polarization and on the geometrical parameters of the system. We also note that for this choice of $a$ and $d$ the quasistatic approximation gives results in perfect agreement with the method using complex $\omega$ and real $k$. We also observe that at both ends of the FBZ and for both polarizations the derivative of $\omega'(k)$ vanish, meaning that [see Eq.~\eqref{Lambda}] the propagation length tends to zero as well.

\begin{figure}[t]
 \includegraphics[width=0.48\textwidth]{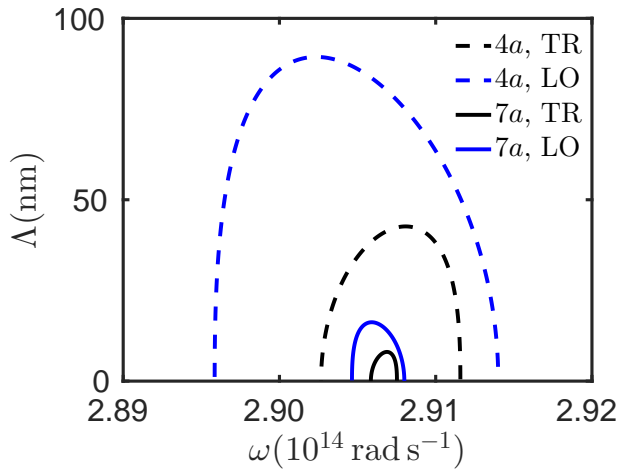}
\caption{Propagation length for transverse (black) and longitudinal (blue) polarizations for a chain of hBN nanoparticles with radius $a=25$ nm. Dashed lines show the results for $d=3a=75\,$nm, whereas solid lines correspond to $d=7a=175\,$nm.}
\label{props}
\end{figure}

\begin{figure}[t]
 \includegraphics[width=0.48\textwidth]{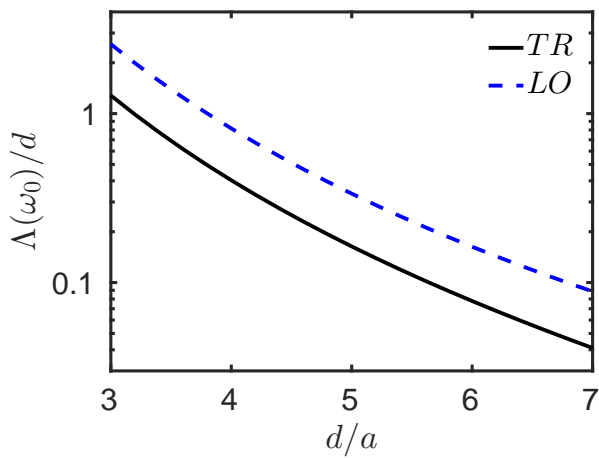}
\caption{Normalized propagation length at the resonance frequency $\omega_0$ as a function of the normalized distance $d/a$ for transverse (black solid line) and longitudinal (blue dashed line) polarization.}
\label{lambdaomega0}
\end{figure}

\subsection{Propagation lengths}

We now focus on the imaginary part $\omega''$ of the frequency, from which we deduce the propagation length $\Lambda$, shown in Fig.~\ref{props} as a function of $\omega$ for both polarizations, and for the two interparticle distances $d=4a=100\,$nm and $d=7a=175\,$nm. We observe two expected main features. First, the range of allowed frequencies decreases when increasing the distance $d$ between the particles. Moreover, the propagation length decreases at any frequency when increasing the distance, which is a clear signature of the reduced coupling between the particles.

To gain more insight into this last point, it is also instructive to study the propagation legnth as a function of the distance $d$. To this end, we need to choose a frequency at which $\Lambda$ can be evaluated for each distance $d$. We focus on the propagation length at the resonance frequency $\omega_0$ and plot in Fig.~\ref{lambdaomega0} the propagation length, renormalized by the distance, as a function of the ratio $d/a$. We observe that the propagation length associated with the longitudinal modes is always larger than the one for transverse modes, and we clearly see the monotonic decrease of $\Lambda$ as a function of $d$ as anticipated before.

\section{Landauer vs Boltzmann approach}\label{SecComparison}

\begin{figure*}[t]
 \includegraphics[width=0.49\linewidth]{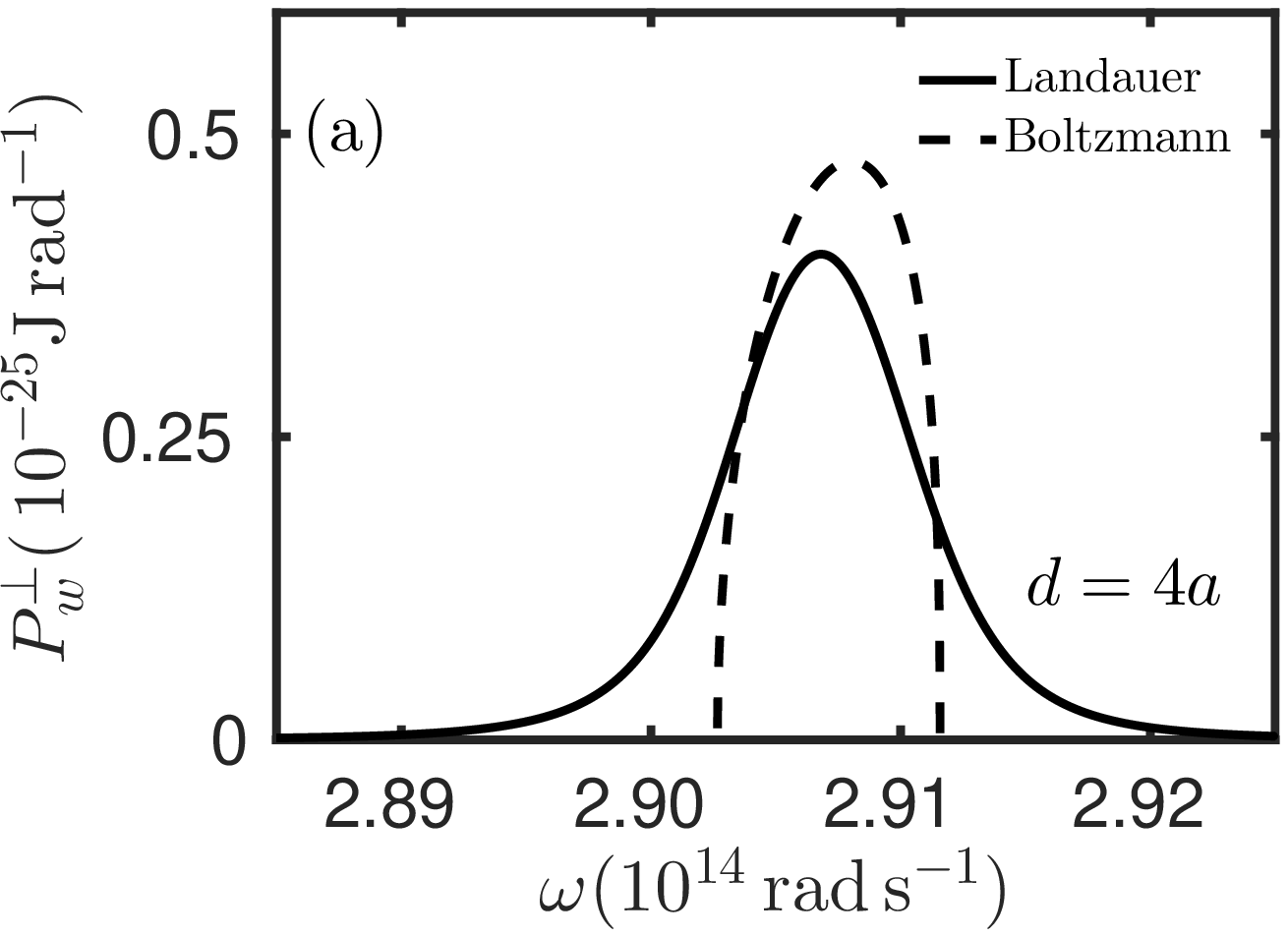}
 \includegraphics[width=0.45\linewidth]{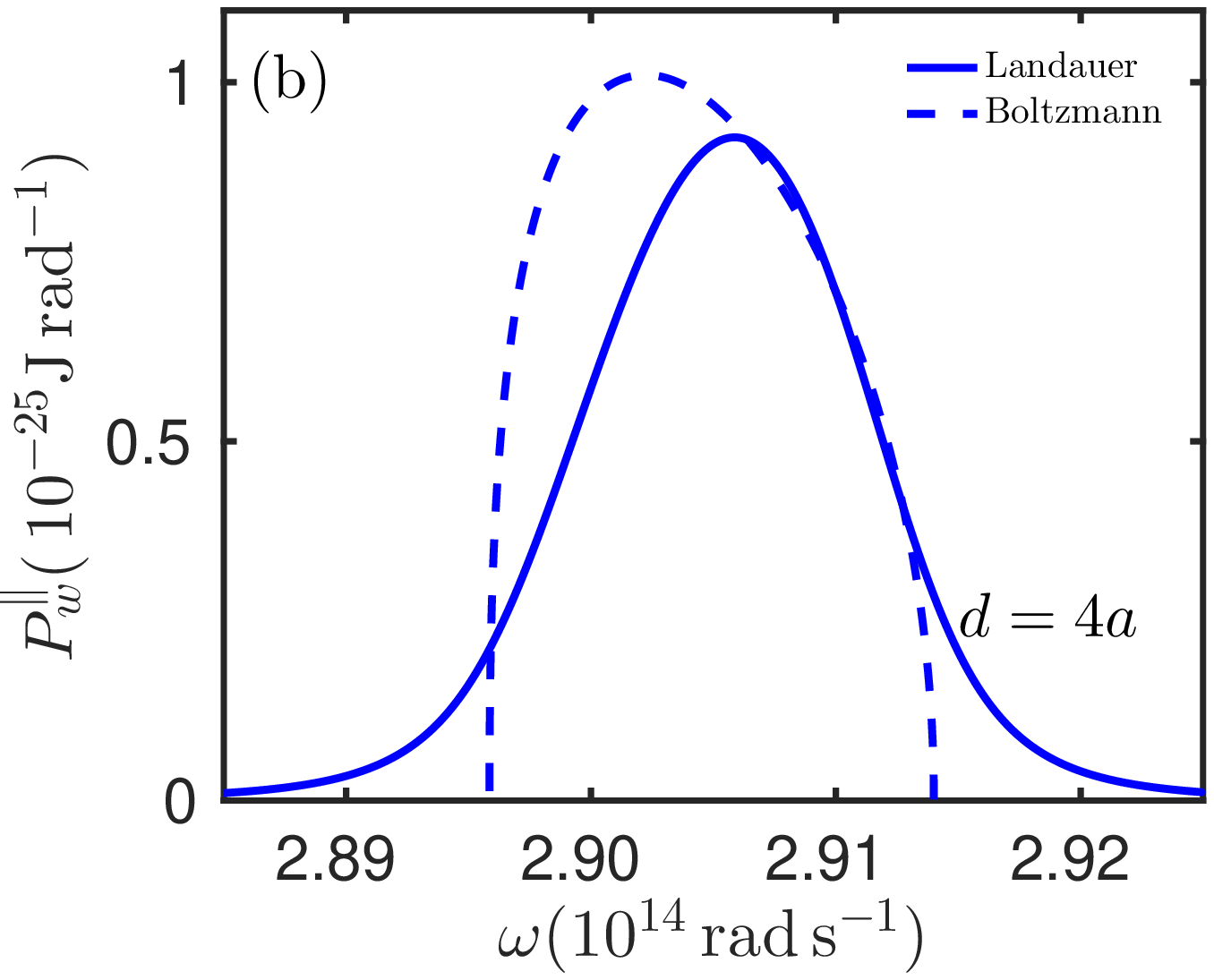}
 \includegraphics[width=0.49\linewidth]{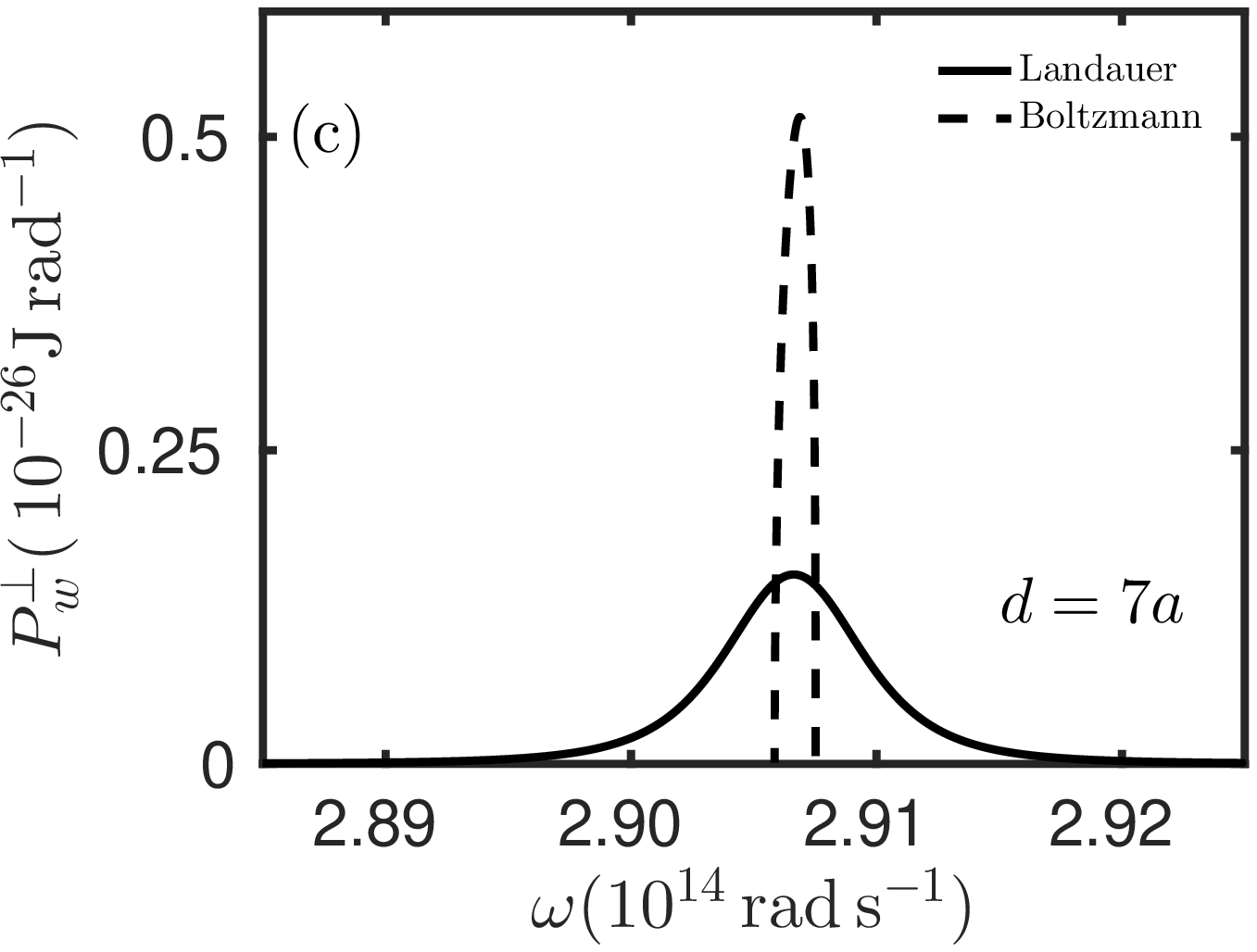}
 \includegraphics[width=0.49\linewidth]{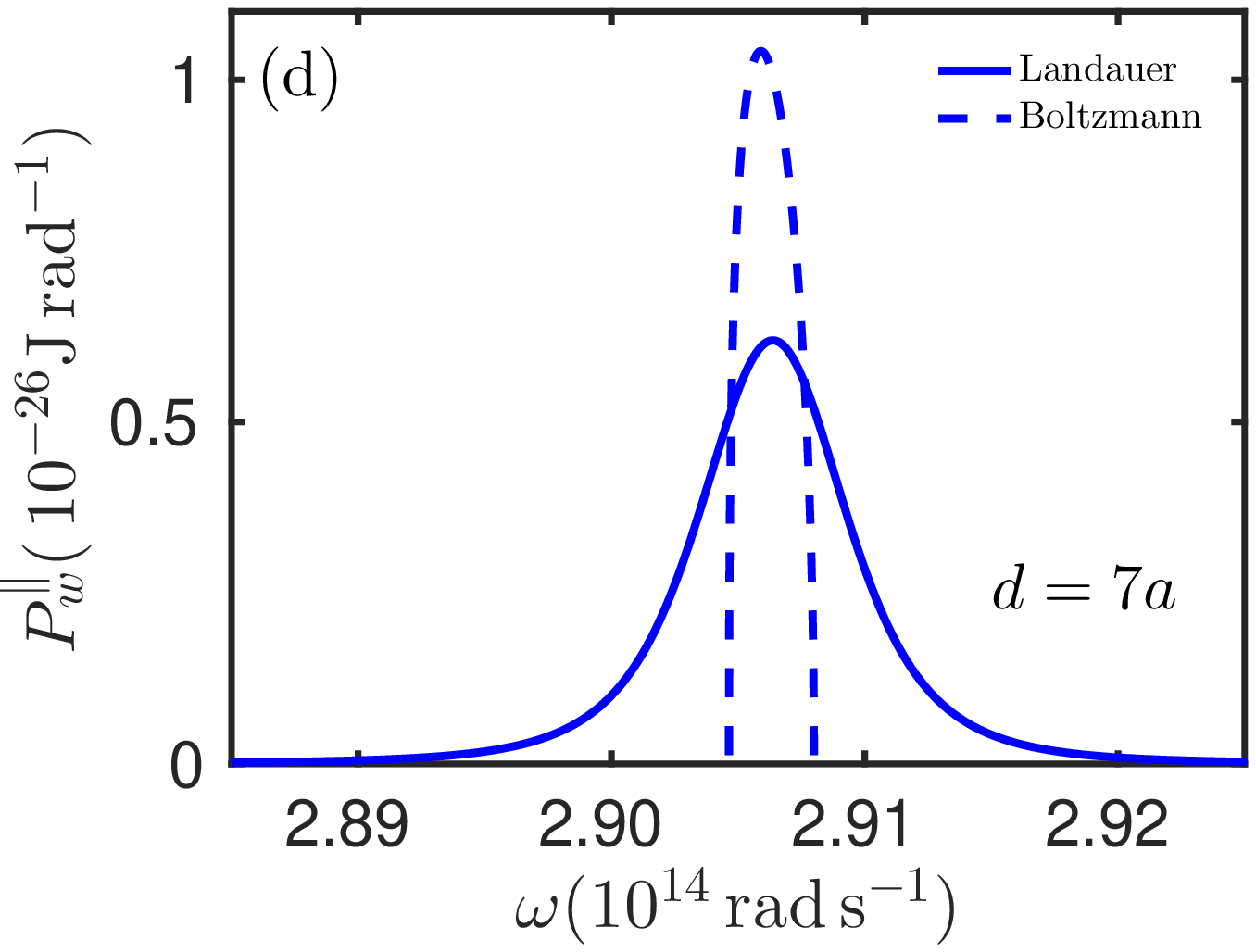}
 \caption{Spectral heat flux decomposed in transverse [(a) and (c)] and longitudinal [(b) and (d)] polarization for $d=4a=100\,$nm [(a) and (b)] and $d=7a=175\,$nm [(c) and (d)]. In all the plots, the solid lines correpond to the Landauer approach, while the dashed lines are associated with the Boltzmann approach.}
\label{n100pspec}
\end{figure*}

We now have all the ingredients needed to calculate the spectral and total flux with the Landauer and Boltzmann approach. As already done in the case of the propagation length, we focus on two different choices of distance, namely $d=4a=100\,$nm and $d=7a=175\,$nm. We show in Fig.~\ref{n100pspec} the spectral fluxes obtained using the two approaches, at both distances and for both polarizations. The first observation is that, while the exact approach gives a flux spectrally defined on the whole frequency spectrum, the Boltzmann approach produces a spectral flux which is non-vanishing only in a specific frequency region, going to zero at the borders of such region. This is of course a direct consequence of the properties of the dispersion relation. We also remark that, even in this region, the Boltzmann approach fails in reproducing the spectral shape of the heat flux, especially at larger distances, and we also note that (in the cases considered here) it either underestimates or overestimates the flux spectrum. This clear spectral discrepancy is the first main result of this work.

\begin{figure}[t]
 \includegraphics[width=0.98\linewidth]{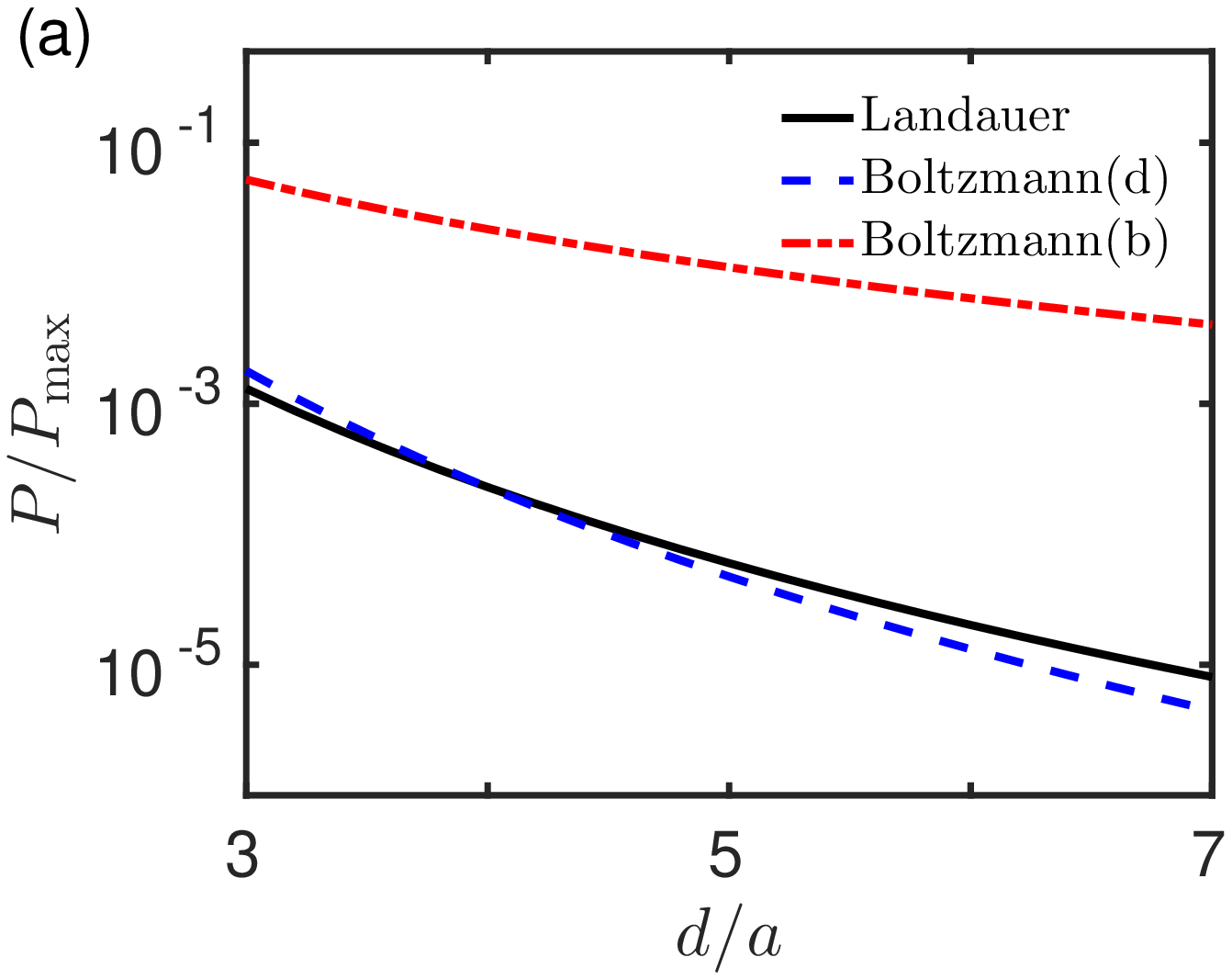}
 \includegraphics[width=0.98\linewidth]{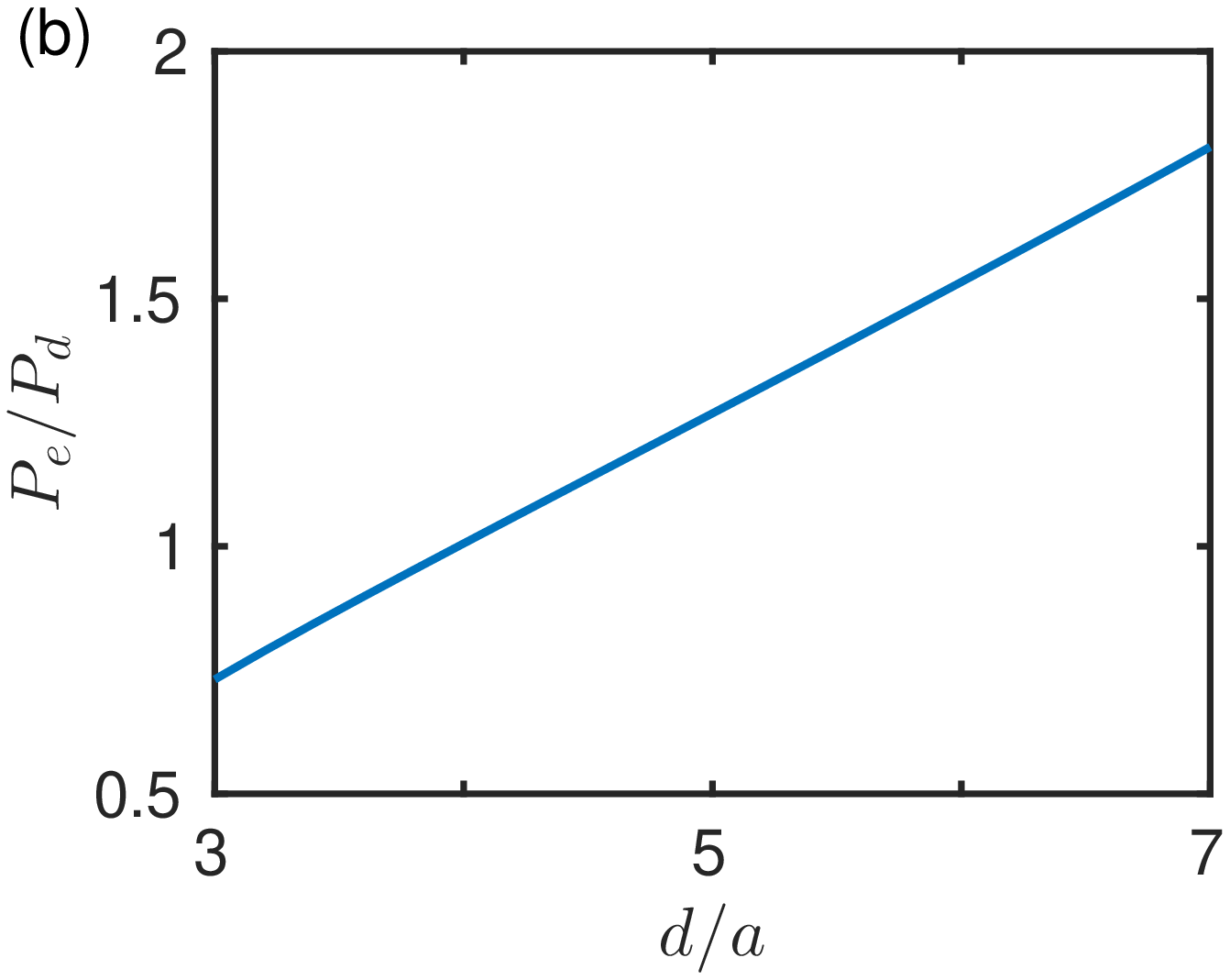}
 \caption{(a) Total flux as a function of distance calculated using Landauer's approach (black solid line), Boltzmann approach in the diffusive regime (blue dashed line) and in the ballistic regime (red dot-dashed line). The fluxes are renormalized with respect to the maximum flux given in Eq.~\eqref{Eq:Pmax}. (b) Ratio between the total flux calculated using Landauer's approach and the Boltzmann diffusive one as a function of the distance.}
\label{totalflux}
\end{figure}

We now turn to the comparison of the total (frequency-integrated) flux as a function of distance, shown in Fig.~\ref{totalflux}. Figure \ref{totalflux}(a) shows three fluxes as a function of the distance: the exact one (calculated using Landauer's approach) and the two corresponding to the diffusive and ballistic approximation within the Boltzmann approach. The three fluxes are divided by the quantity
\begin{equation}
 P_{\rm max}=3\int_{\omega_{\rm TO}}^{\omega_{\rm LO}} \frac{d\omega}{2\pi} \hbar \omega\,n_{1N}(\omega),
\label{Eq:Pmax}
\end{equation}
representing the largest possible flux in the Boltzmann approach, since each allowed mode is assumed to contribute with transmission coefficient equal to 1 and $\omega_{\rm TO}$ and $\omega_{\rm LO}$ are the smallest and largest allowed values for $\omega_\text{min}$ and $\omega_\text{max}$, respectively. We first observe in Fig.~\ref{totalflux}(a) that both limiting cases in the Boltzmann approach give results far (more than one order of magnitude) from this maximum flux. Concerning the comparison between these two results and the exact one, we start by noting that the ballistic regime is in complete disagreement with the correct result, with a ratio going as high as two orders of magnitude. This is first due to the fact that for most distances we are clearly in the diffusive regime, as can be seen in Fig.~\ref{lambdaomega0}. On the other hand, even for $d = 3a$ where the transition towards the ballistic regime occurs, the ballistic result is overestimating the heat flux by more than one order of magnitude. Concerning the results from the Boltzmann approach in the diffusive regime, we can clearly see that while it grasps the overall behavior of the flux, there is a significant discrepancy which increases when increasing the distance. Quantitavely speaking, as can be seen in Fig.~\ref{totalflux}(b), for the smallest distance of $d=3a=75\,$nm the error is approximately 10\% and monotonically increases with the distance up to an error of a factor of about 2 for a relatively large distance of $d = 7a = 175$nm.

It is interesting to observe that the ratio between the exact and the Boltzmann diffusive powers increases monotonically as a function of $d$. This is the result of a different power-law scaling of the two quantities as a function of the distance $d$. As a matter of fact, we have verified that in the Landauer approach the power scales as $d^{-6}$: this is indeed a signature of the reduced $N$-body coupling for larger $d$, since we recover in diluted chains the behavior of the power exchange in the dipole-dipole configuration. Concerning the Boltzmann results, we have numerically observed that in the ballistic regime the power scales as $d^{-3}$. From Eq.~\eqref{Eq:ballisticflux}, we clearly see that this results only depends on the amplitude of the interval of allowed frequencies $\omega_\text{max}-\omega_\text{min}$, which shrinks when increasing $d$ as shown in Sec.~\ref{SecModes}B (see Fig.~\ref{props}). Concerning the diffusive approach, we see from Eq.~\eqref{DiffFin} that it depends also on the propagation lengths in the two polarizations. We have verified that these scale as $d^{-4}$. As a result, the scaling of the exchanged power in the diffusive regime is $d^{-7}$, which gives the linear ratio shown in Fig.~\ref{totalflux}(b).

\section{Discussion}\label{SecDiscussion}

We have presented a comparison between an exact Landauer approach and an approximative kinetic Boltzmann approach for the evaluation of the radiative heat transfer along a linear chain of nanoparticles. In the scenario under scrutiny, all the particles were made of the same material (hBN), having a resonance in the infrared region of the spectrum. But even in this simple scheme, we have shown that the kinetic approach completely fails to predict both the heat flux spectra and the exact value of heat flux both in the diffusive and the ballistic regimes of heat transport. Before concluding, we would like to anticipate a generalization of our present work to more complex systems than simple 1D nanoparticle chains. The extension of the kinetic approach to more complex 2D and 3D structure is indeed a more delicate topic. If, on the one hand, the calculation of the dispersion relation can become very tricky, on the other hand the radiative heat transfer could imply unexpected collective $N$-body effects which are fully taken into account by Landauer's approach and could be missed by the kinetic one. Moreover, as we have shown an intrinsic limitation of the kinetic approach is that it only takes into account the contribution of resonant modes so that a huge number of potential heat carriers are simply neglected.

Beside this geometric aspect the materials involved in the system also can make the Boltzmann approach unpractical. Indeed for simple chains of particles made of polar materials the surface phonon-polaritons give the dominant contribution to the radiative heat transfer if their resonance frequency is well within the Planck window imposed by the temperatures present in the system. However, the situation is expected to radically change when the resonant modes are located outside of this spectral range. Hence at ambient temperature if those resonances are in the ultraviolet range (as for metals) the surface resonance is no longer supposed to give the main contribution to the flux. While this would be automatically accounted for in the exact calculation, giving the contribution at any frequency, it would be completely ignored by the kinetic approach, giving only the contribution at frequencies around the resonance.
 
The use of the kinetic approach can also lead to strong deviations with respect to the exact theory if the system supports a continum of modes which superimpose to the presence of a surface mode inside the Planck window. This situation can occur, for instance, in hyperbolic materials, uniaxial anisotropic media made with a periodic combination of dielectric and metallic components. For these materials it has been shown that the main contribution to the flux can in certain circumstances mainly come from the participation of a continum of modes also called hyperbolic modes~\cite{Hyper} and not from the presence of surfaces modes. In those situations, the kinetic approach should underestimate the value of heat flux.

Finally, for the same fundamental reasons as discussed in the present work to evaluate radiative heat exchanges in many-body systems strong deviations are expected to occur in others branches in physics. Hence the estimate of the phonon thermal conduction~\cite{Rego,Yamamoto} as well as the thermoelectric transport coefficients of nanostructured materials~\cite{Jeong} could be strongly impacted by the use of the Boltzmann or Landauer theoretical frameworks. 

In conclusion, we have shown that the Boltzmann approach is a very limited tool for the description of radiative heat transfer in a system made of nanoparticles. In both the diffusive and ballistic approximations, it completely fails in describing spectrally the heat exchange, and also the total heat flux differs by a factor which was shown to be close to 2 in a very simple configuration. This deviation from the exact results is expected to increase significantly in numerous systems. Hence, the application of the kinetic theory is extremly delicate or even unpractical for the proper investigation of radiative heat transfer in many-body systems.

\acknowledgments

The authors acknowledge financial support by the DAAD and Partenariat Hubert Curien Procope Program (project 57388963).


\begin{thebibliography}{99}
\bibitem{Kirchoff} G. Kirchhoff, \emph{On the Relation between the Radiating and Absorbing Powers of Different Bodies for Light and Heat}, Philos. Mag. Ser. 5 \textbf{20}, 1 (1860).
\bibitem{Planck} M. Planck, \emph{The Theory of Heat Radiation} (Dover, New York, 1991).
\bibitem{Polder} D. Polder and M. Van Hove, \emph{Theory of Radiative Heat Transfer between Closely Spaced Bodies}, Phys. Rev. B \textbf{4}, 3303 (1971).
\bibitem{1} P. Ben-Abdallah, \textit{Heat transfer through near-field interactions in nanofluids}, Appl. Phys. Lett. \textbf{89}, 113117 (2006).
\bibitem{2} P. Ben-Abdallah, K. Joulain, J. Drevillon, and C. Le Goff, \textit{Heat transport through plasmonic interactions in closely spaced metallic nanoparticle chains}, Phys. Rev. B \textbf{77}, 075417 (2008).
\bibitem{Rytov}S.~M. Rytov, Y.~A. Kravtsov, V.~I. Tatarski, \emph{Principles of Statistical Radiophysics} (Springer, Berlin, 1987).
\bibitem{3} J. Ordonez-Miranda, L. Tranchant, S. Gluchko, and S. Volz, \textit{Energy transport of surface phonon polaritons propagating along a chain of spheroidal nanoparticles}, Phys. Rev. B \textbf{92}, 115409 (2015).
\bibitem{5} J. Ordonez-Miranda, L. Tranchant, K. Joulain, Y. Ezzahri, J. Drevillon, and S. Volz, \textit{Thermal energy transport in a surface phonon-polariton crystal}, Phys. Rev. B \textbf{93}, 035428 (2016).
\bibitem{6} J. Ordonez-Miranda, L. Tranchant, B. Kim, Y. Chalopin, T. Antoni, and S. Volz, \textit{Quantized Thermal Conductance of Nanowires at Room Temperature Due to Zenneck Surface-Phonon Polaritons}, Phys. Rev. Lett. \textbf{112}, 055901 (2014).
\bibitem{4} F.~V. Ramirez and A.~J.~H. McGaughey, \textit{Plasmonic thermal transport in graphene nanodisk waveguides}, Phys. Rev. B \textbf{96}, 165428 (2017).
\bibitem{9} P. Ben-Abdallah, S.-A. Biehs, and K. Joulain, \textit{Many-body radiative heat transfer theory}, Phys. Rev. Lett. \textbf{107}, 114301 (2011).
\bibitem{10} R. Messina, M. Tschikin, S.-A. Biehs, and P. Ben-Abdallah, \textit{Fluctuation-electrodynamic theory and dynamics of heat transfer in systems of multiple dipoles}, Phys. Rev. B \textbf{88}, 104307 (2013).
\bibitem{8} R. Incardone, T. Emig, and M. Kr\"{u}ger, \textit{Heat transfer between anisotropic nanoparticles: Enhancement and switching}, Europhys. Lett. \textbf{106}, 41001 (2014).
\bibitem{Edalatpour2014} S. Edalatpour and M. Francoeur, \textit{The Thermal Discrete Dipole Approximation (T-DDA) for near-field radiative heat transfer simulations in three-dimensional arbitrary geometries}, J. Quant. Spec. Rad. Trans. \textbf{133}, 364 (2014).
\bibitem{11} M. Nikbakht, \textit{Radiative heat transfer in anisotropic many-body systems: Tuning and enhancement}, J. Appl. Phys. \textbf{116}, 094307 (2014).
\bibitem{12} Y. Wang and J. Wu, \textit{Radiative heat transfer between nanoparticles enhanced by intermediate particle}, AIP Advances \textbf{6}, 025104 (2016).
\bibitem{EkerothEtAl2017} R.~M. Abraham Ekeroth, A. Garc\'ia-Mart\'in, and J.~C. Cuevas, \textit{Thermal discrete dipole approximation for the description of thermal emission and radiative heat transfer of magneto-optical systems}, Phys. Rev. B \textbf{95}, 235428 (2017).
\bibitem{MA14} R. Messina and M. Antezza, \emph{Three-body radiative heat transfer and Casimir-Lifshitz force out of thermal equilibrium for arbitrary bodies}, Phys. Rev. A \textbf{89}, 052104, (2014).
\bibitem{14} M. Nikbakht, \textit{Radiative heat transfer in fractal structures}, Phys. Rev. B \textbf{96}, 125436 (2017).
\bibitem{16} I. Latella, P. Ben-Abdallah, S.-A. Biehs, M. Antezza, and R. Messina, \textit{Radiative heat transfer and nonequilibrium Casimir-Lifshitz force in many-body systems with planar geometry}, Phys. Rev. B \textbf{95}, 205404 (2017).
\bibitem{13} P. Ben-Abdallah, R. Messina, S.-A. Biehs, M. Tschikin, K. Joulain, and C. Henkel, \textit{Heat superdiffusion in plasmonic nanostructure networks}, Phys. Rev. Lett. \textbf{111}, 174301 (2013).
\bibitem{17} I. Latella, S.-A. Biehs, R. Messina, A.~W. Rodriguez, and P. Ben-Abdallah, \textit{Ballistic near-field heat transport in dense many-body systems}, Phys. Rev. B \textbf{97}, 035423 (2018).
\bibitem{MessinaPRL13} R. Messina, M. Antezza, and P. Ben-Abdallah, \emph{Three-Body Amplification of Photon Heat Tunneling}, Phys. Rev. Lett. \textbf{109}, 244302 (2012).
\bibitem{PBA16} P. Ben-Abdallah, \emph{Photon Thermal Hall Effect}, Phys. Rev. Lett. \textbf{116}, 084301 (2016).
\bibitem{Transistor} P. Ben-Abdallah and S.-A. Biehs, \emph{Near-Field Thermal Transistor}, Phys. Rev. Lett. \textbf{112}, 044301 (2014).
\bibitem{Memory} V. Kubytskyi, S.-A. Biehs, and P. Ben-Abdallah, \emph{Radiative Bistability and Thermal Memory}, Phys. Rev. Lett. \textbf{113}, 074301 (2014).
\bibitem{Boolean} P. Ben-Abdallah and S.-A. Biehs, \emph{Towards Boolean operations with thermal photons}, Phys. Rev. B \textbf{94}, 241401(R) (2016).
\bibitem{NarayanaswamyApplPhysLett03} A. Narayanaswamy and G. Chen, \emph{Surface modes for near field thermophotovoltaics}, Appl. Phys. Lett. \textbf{82}, 3544 (2003).
\bibitem{spitzer} W.~G. Spitzer, D. Kleinman, and D. Walsh, \textit{Infrared Properties of Hexagonal Silicon Carbide}, Phys. Rev. \textbf{113}, 127 (1959).
\bibitem{abajo} A. Manjavacas and F.~J. Garcia de Abajo, \textit{Radiative heat transfer between neighboring particles}, Phys. Rev. B \textbf{86}, 075466 (2012).
\bibitem{haken} H. Haken, \textit{Quantenfeldtheorie des Festk\"{o}rpers}, second edition (B.~G. Teubner, Stuttgart, 1993).
\bibitem{ford} W.~H. Weber and G.~W. Ford, \textit{Propagation of optical excitations by dipolar interactions in metal nanoparitcle chains}, Phys. Rev. B \textbf{70}, 125429 (2004).
\bibitem{Hyper}S.-A. Biehs, M. Tschikin, and P. Ben-Abdallah, \emph{Towards a black body for near-field thermal radiation}, Phys. Rev. Lett. \textbf{109}, 104301 (2012).
\bibitem{Rego} L. G.C. Rego and G. Kirczenow, Phys. Rev. Lett. \textbf{81}, 232 (1998).
\bibitem{Yamamoto} T. Yamamoto and K. Watanabe, Phys. Rev. Lett. \textbf{96}, 255503 (2006).
\bibitem{Jeong} C. Jeong, R. Kim, M. Luisier, S. Datta and M. Lundstrom, J. Appl. Phys., \textbf{107}, 023707 (2010).
\end{thebibliography}
\end{document}